# Superconducting and Structural Features of Tl-1223

R. Shipra,[1,2] J.C. Idrobo,[3,1] and A.S. Sefat [2,*]

[1]*Department of Physics and Astronomy, Vanderbilt University, Nashville, Tennessee 37235, USA*

[2]*Materials Science and Technology Division, Oak Ridge National Laboratory, Oak Ridge, Tennessee 37831, USA*

[3]*Center for Nanophase Materials Sciences, Oak Ridge National Laboratory, Oak Ridge, Tennessee 37831, USA*

* Corresponding author: *sefata@ornl.gov*

**Abstract**

This study provides an account of the bulk preparation of $TlBa_2Ca_2Cu_3O_{9-\delta}$ (Tl-1223) superconductor at ambient pressure, and the $T_c$ features under thermal-annealing conditions. The 'as-prepared' Tl-1223 ($T_c$ =106 K) presents a significantly higher $T_c$ = 125 K after annealing the polycrystalline material in either flowing Ar+4% $H_2$, or $N_2$ gases. In order to understand the fundamental reasons for a particular $T_c$, we refined the average bulk structures using powder X-ray diffraction data. Although Ar+4%$H_2$ annealed Tl-1223 shows an increased '$c$' lattice parameter, it shrinks by 0.03% (approximately unchanged) upon $N_2$ anneal. Due to such indeterminate conclusions on the average structural changes, local structures were investigated at using aberration-corrected scanning-transmission electron microscopy technique. Similar compositional changes in the atomic arrangements of both annealed-samples of Tl-1223 were detected in which the plane containing Ca atomic layer gives a non-uniform contrast, due to substitution of some heavier Tl. In this report, extensive bulk properties are summarized through temperature-dependent resistivity, and shielding and Meissner fractions of magnetic susceptibility results; the bulk and local structures are investigated to correlate to properties.

**Introduction**

Thermal annealing, in the presence or absence of vacuum, oxygen or other reactive and neutral gases, has become a conventional practice in studies of high-temperature cuprate superconductors (HTS).[1,2] Such a method can lead to changes in oxygen stoichiometry and carrier concentration, which in-turn plays an important role in getting an optimal critical temperature ($T_c$) in these cuprates.[1,3-5,6,7] A direct relation between the $T_c$ and the oxygen content has been successfully established in Hg-based HTS, but Bi- and Tl-based layered cuprates present much more complex dependence.[6,8,9] Such complexities are reported as chemical inhomogeneity, disorder, and related inter-site cation exchanges, highly dependent on the starting composition and preparative conditions.[6,8] Thallium based cuprate superconductors are described by the general formula of $Tl_mBa_2Ca_{n-1}Cu_nO_{2n+m+2}$ abbreviated as Tl-$m2(n-1)n$, where $m$ = 1 or 2 and $n$ = 1, 2, 3 or 4. The Tl-based cuprates have an advantage over Bi-based superconductors, since $m$ = 1 has smaller structural anisotropy with a low lattice $c/a$ ratio, which may be prone to more isotropic transport properties, comparable to those of Hg-based HTS.[7]

Among the members of Tl-based series, Tl-1223 (or $TlBa_2Ca_2Cu_3O_{9-\delta}$) is one of the preferable candidate superconductors for fundamental and applied research because its very high $T_c$ reaching values of 125 K, and its efficient flux pinning properties.[10-12] The Tl-1223 phase, though thermodynamically stable, is difficult to form at ambient pressure (AP) compared with other members of Tl-based cuprates.[11,13] The substituted phases of Tl-1223 have been studied a lot for structural domain stabilization, in comparison to pure Tl-1223, through substitution studies of $Sr^{2+}$ for $Ba^{2+}$, and $Pb^{+4}$ or $Bi^{+3}$ for $Tl^{+3}$.[4,12,14] Besides phase stability, the electric-current carrying ability of Tl-1223 superconductor also improves upon chemical substitution.[10] Earlier, $T_c$ values of ~ 132-133 K were achieved in high-pressure (HP) synthesized Tl-1223, after annealing in nitrogen. Such $T_c$ values are easily comparable to Hg-1223 HTS.[3,4,15,16] The 'a' and 'c' refined lattice parameters of the HP Tl-1223 phase were found to be similar to that of AP phase with $T_c$ ~ 126 K. On the atomic scale, the structural layer sequence in Tl-1223 is $(TlO)^{+1}$-$(BaO)^{0}$-$(CuO_2)^{-2}$-$Ca^{+2}$-$(CuO_2)^{-2}$-$Ca^{+2}$-$(CuO_2)^{-2}$-$(BaO)^{0}$-$(TlO)^{+1}$, where Cu is adjacent to the BaO plane forming a square pyramid coordination with oxygen, while the $CuO_2$ planes, sandwiched between Ca layers form a square planar coordination. Structural refinement models based on neutron and X-ray diffraction suggest the possibility of the presence of two Tl atomic positions for different oxidation states in the charge reservoir layer (consisting of a single Tl-O layer) of Tl-1223.[4,15] One of the Tl ions, $Tl^{+1}$, sits on the four fold symmetry site (0,0,0), while $Tl^{+3}$ is usually located off-site at (0.10, 0, 0).[4,15] Their distribution is statistical and the relative amounts of each may depend on the materials' synthesis and annealing conditions. The ratio of $Tl^{+3}$:$Tl^{+1}$ was found to be as high as 5:1 in Tl-1223 prepared at AP, increasing to 7:1 in HP-prepared sample.[15,17] It has also been observed that increasing oxygen occupancy enhances the amount of $Tl^{+3}$, which has a smaller size compared to $Tl^{+1}$, resulting in a Tl-O contracted layer along the $c$-axis during oxidation.[15] According to Morosin *et. al.*, this contraction attracts the neighboring Ba towards the Tl-O layer while at the same time repels O towards the $CuO_2$ plane. The contraction of the Tl-O layer along the $c$-axis results in decrease of the Cu-$O_{apical}$ length and an increment of the positive charge on the Cu ions.[15,18] Though Cu-$O_{apical}$ bond length may not have a direct influence in determining $T_c$ in multilayered hole-doped cuprates, experimental and theoretical evidence suggest its role with relation to the number of holes that affect $T_c$; this explanation has been presented in HTS $HgBa_2Ca_2Cu_3O_{8+\delta}$, $La_2CuO_{4+\delta}$ and $YBa_2Cu_3O_{7-\delta}$.[3,4,6,18] In fact, both Tl-1223 and Hg-1223 reach almost the same Cu-$O_{apical}$ length for maximum $T_c$. For Tl-1223, neutron diffraction gives Cu-$O_{apical}$ = 2.747 Å for the highest $T_c$ (132 K). For Hg-1223, the optimum Cu-$O_{apical}$ =2.741 Å for the highest $T_c$.[19] Generally, $T_c$-structure relations are still quite inconclusive. For example, although there is monotonic variation of lattice parameters with $T_c$ for HP-prepared and (argon) annealed Tl-1223,[3] different values of $T_c$ can be obtained for the same lattice structural values.[4] It is worth mentioning that charge doping in Tl-based HTS is also influenced by various cationic interchanges such as $Tl^{3+}$ substituting $Ca^{2+}$ in between the $CuO_2$ planes. As suggested by Morosin *et. al.*, such substitutions may be around 5% in Tl-1223 AP-prepared, in contrast to 0.3 % in the HP-prepared sample with the highest $T_c$.[4,15] Various structural polymorphs simultaneously occur in Tl-based cuprates as lower members ($n < 3$) form at an early stage of growth before converting to higher phase ($n \geq 3$) especially in the case of Tl-1223 and Tl-2223.[20-22] Slight variations in the bulk composition, though not detectable in the X-ray diffraction patterns (slightly broaden Bragg peak; oxygen insensitivity), may coexist with desired composition and affect the deterioration or broadening of the superconducting transition.[20] The structural arrangement of atoms on the local scale, vacancies, defects, and disorder can be seen in high-resolution transmission-electron microscopy (TEM). Presence of vacancies, stacking faults and intergrowths have been confirmed in Tl-1223, similar to other Tl-based HTS.[20,23] Hopfinger *et. al.* found several non-equilibrium intergrowth phases between Tl-1223 and Tl-2223 in fluorine substituted Tl-F-1223 samples, which may be candidates for high-current applications.[20]

In this study, we report a straight-forward method to synthesize Tl-1223 HTS at ambient pressures, and then thermally anneal in various conditions to check for changes in superconducting properties. We summarize the bulk behavior of temperature-dependent resistivity, heat capacity, and magnetic susceptibility, and estimate the critical current density in magnetization. We correlate variations of $T_c$ with local structures obtained with TEM and bulk structure obtained with X-ray diffraction, in order to shed some light on reasons for the change in $T_c$.

**Experimental**

For synthesizing Tl-1223, we used the usual precursor route in which first a precursor is prepared having a molar composition equivalent to $Ba_2Ca_2Cu_3O_7$ and then a suitable amount of $Tl_2O_3$ is added to complete the formation of the desired compound. For this, $BaCO_3$, $CaCO_3$ and $CuO$, in the cationic ratio of 2:2:3 respectively, were ground and calcined in air three times at subsequent temperatures of 880, 890 and 900 °C, respectively; each calcination was done for 24 hrs. The final mixture was pelletized at a pressure of 6000 tons/m$^2$ and sintered at 910 °C for 48 hrs. The X-ray diffraction (XRD) pattern showed the final mixture consisting of $BaCuO_2$, $Ca_2CuO_3$ and $CuO$, to which $Tl_2O_3$ was added. For synthesizing Tl-1223 in 'open tubes', a slight excess amount of $Tl_2O_3$ was required, which is very obvious when one considers its volatile nature. The best optimized ratio of $Tl_2O_3$ to precursor for preparing Tl-1223 was between 0.9:1 and 1:1. Sample 1 was prepared using the ratio 0.9:1 (contained $BaCuO_2$ impurities), while sample 2 was prepared using the ratio 1:1 (contained Tl-1212 impurity phase). We further tried to reduce the amount of $Tl_2O_3$ as well as the sintering time, but found large amounts of $BaCuO_2$ impurities. The mixed $Tl_2O_3$ and precursor powder were ground well and then pressed by applying a pressure of 1500 tons/m$^2$ into ¼ inch pellets each weighing between 0.5 - 0.75 grams. Each pellet was individually wrapped in gold foil and then synthesized to Tl-1223 phase at 910 °C for 3 hrs under flowing $O_2$. To study the variations in structure-property relation, and $T_c$ with annealing, sample 1 was first annealed in a sealed quartz tube under 1 atm. $O_2$ (at 720 °C for 48 hrs) and then annealed three times successively in a mixture of flowing Ar+4%$H_2$ at 250 °C (1 hour each time). Also, sample 2 was first annealed in oxygen, then powdered and annealed under flowing $N_2$ at 675 °C for 1.5 hours. Sample 2 was further process by annealing at 675 °C (1 hour). All the above mentioned synthesis conditions were well optimized, and are highly reproducible, and unique, as excess $Tl_2O_3$ (nominal composition of the initial reactants is equivalent to that of Tl-2223) was used in the single step synthesis under flowing $O_2$. The optimized synthesis condition also provides a good example for the gradual conversion of Tl-2223 to Tl-1223 as it has already been suggested earlier.[21,24]

The phase purity is more than 90% in all the samples that are studied here. The XRD patterns were obtained using X'pert PRO MPD powder diffractometer in the 5-90° 2θ range. The refined cell parameters as well as the bond lengths were determined from Lebail refinements/matching of the experimental data using (reference code: 01-081-0044) HighScore Plus software. Zero shift, unit-cell dimensions and profile values are the parameters refined. The $T_c$ and $J_c$ values were estimated by performing magnetization measurements performed using SQUID magnetometer by Quantum Design (QD). The zero-field-cooled (ZFC) data were collected while increasing the temperature from 5 K to 150 K after applying the field at 5 K; the field-cooled (FC) data were obtained while cooling in the same field from 150 K. For some samples we correlated the onset of diamagnetic transition with that of the resistive transitions by doing four-probe temperature-dependent transport measurements in a QD's Physical Property Measurement System (PPMS). Aberration-corrected scanning tunneling electron microscopy (STEM) was performed on samples 1, 1-4 and 2-3 to locate changes in the atomic arrangements before and after annealing. STEM imaging were performed in a Nion UltraSTEM 100, which has a cold field-emission electron source and that can correct 3rd and 5th-order aberrations.[25] The microscope was operated at 100 kV accelerating voltage, using a 30 mrad semi-convergence angle, and 80-200 mrad semi-collection angles, for the annular dark field detector.

**Results and discussion**

Excess $Tl_2O_3$ was used to compensate for Tl loss and more based on earlier reports that suggest the formation of Tl-1223 progresses through de-intercalation of Tl from Tl-2223.[24] It is documented in the literature that nucleation of Tl-1223 phase at ambient pressure starts to occur at temperatures above 905 °C.[21] It has also been observed that Tl-1223 decomposes completely at around 940 °C, which narrows the reaction temperatures.[21] The sample synthesized at 915°C contained large amounts of BaCuO$_2$, hence we restricted ourselves to a sintering temperature of 910°C. After synthesizing many samples, it was decided that any decrease in the sintering temperature results in the formation of other members of Tl-based cuprates consisting of both single and double Tl-O layers.

Figure 1 compares the XRD patterns of the as-prepared samples and those annealed. All the peaks are well resolved and indicate good crystallinity of the material. Small amounts of Ba$_{44}$Cu$_{48}$O$_{100}$ (BCO; ~ 5 % by weight, marked with *) and CaO (2.4%, not shown in the figure) are present as the only impurity phase in the as-prepared sample 1. The impurity concentrations are only slightly affected after annealing in Ar+4%H$_2$ (in sample 1-4). Sample 2 contains Tl-1212 (7.6%) which disappears while Ba$_{44}$Cu$_{48}$O$_{100}$ (10.6%) appears after the sample (2-1) was ground and annealed under flowing N$_2$. CaO is present in very small amounts (1.7% in sample 2 and 2.4% in sample 2-3). Detailed structural parameters derived from the structural refinement of the XRD patterns of the prepared samples are given in Table 1, along with their synthesis conditions. The as-prepared sample 2 contains Tl-1212 as secondary phase, which disappears, while BCO phase appears after N$_2$ annealing. Sample 1, however, does not contain any Tl-1212 while BCO is present in considerable amounts. Sample 2 was prepared using more Tl$_2$O$_3$ and comparatively low amounts of CaO in the XRD pattern may explain the presence of the Tl-1212 phase. Phase diagrams suggest that the separation of Tl-1212 mainly occur during the synthesis of Sr- based analogue of Tl-1223 superconductor.[11] A decrease in the amounts of BaCuO$_2$, Tl-1212 and CaO impurities was also observed after long time O$_2$ anneal in both samples while even a short-time annealing under reduced atmosphere increased the impurities as evident from Table 1. O$_2$ annealing at 720°C, which is just above the melting temperature of Tl$_2$O$_3$, was performed to homogeneously distribute Tl and O in the samples 1 and 2. We observe a decrease in the amounts of BaCuO$_2$, Tl-1212 and CaO. It was also found that more than one time such annealing was not useful, though a diamagnetic transition temperature of ~ 123 K can be achieved after 4 such heating. We found this method to be time-consuming, hence annealed samples under either flowing Ar+4%H$_2$, or N$_2$, to optimize $T_c$ similarly to those reported.[2,26] For sample 1, we observe a systematic increase in both '$a$' and '$c$' refined lattice parameters from XRD, and $T_c$, with annealing. For sample 2, for the same rise in $T_c$, we observed that the '$a$' increases, while the '$c$' lattice parameter contracts. Similar response on temperature for the lattice parameters was reported on N$_2$-annealed HP synthesized samples.[26] Expansion of '$c$' lattice parameter upon Ar+4%H$_2$ reduction, however presents an opposite behavior, which is then similar to that reported in Ar annealed samples.[3] For the same $T_c$ (~ 125 K), the '$c$' lattice parameter of sample 1-4 is 15.9261(4) Å, whereas for sample 2-3 it refines as 15.8755(4) Å. In multilayered cuprates, '$c$' axis variations are mainly linked to the elemental arrangement, vacancies, or substitutions, while variations in '$a$' are mainly linked to doping (substitutions, carrier concentration).[15] In this respect, if we consider same doping levels then similar $T_c$ values for both samples can be understood where the '$a$' lattice as well as the Cu and planar oxygen (Cu-O$_{planar}$) bond length expand to attain almost identical values. Also, differences in the '*c*' lattice parameter and Cu apical oxygen (Cu-O$_{apical}$) bond length indicate substantial dissimilarities between the elemental arrangements, as well as cationic interchanges in the non-Cu-O layers available for the mediation of charge carriers to the Cu-O planes. As discussed in the introduction, to obtain a certain $T_c$, the structure relation in these compounds are inconclusive, even in highly-studied HP synthesized samples.[3,4] Thalium's volatile nature and its tendency to change oxidation states (+1 or 3), along with the statistical distribution of various elemental defects, complicates the control and understanding of $T_c$, but that reduction annealing procedure always improves $T_c$.

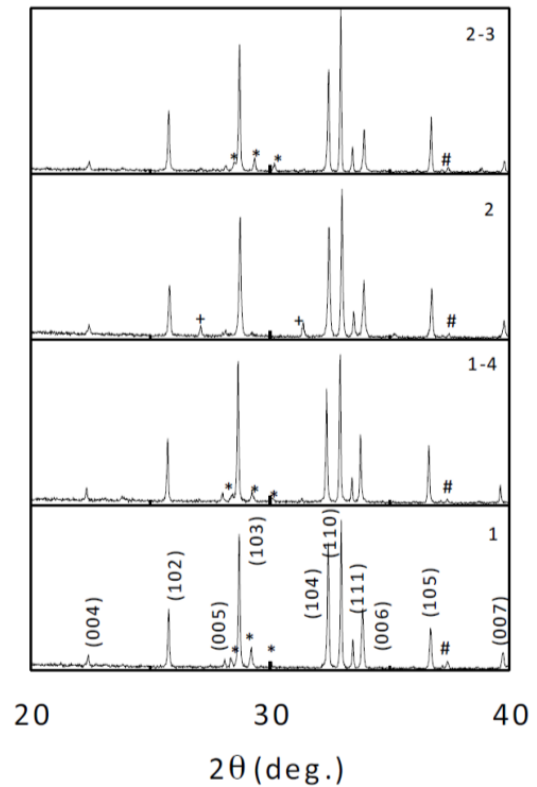

**Figure 1.** X-ray diffraction patterns of the as-prepared and annealed Tl-1223 samples (see table 1 for sample referencing). Ba$_{44}$Cu$_{48}$O$_{100}$, CaO and Tl-1212 impurities are denoted as *, #, and + symbols, respectively. All peaks are normalized to the (110) peak.

Table 1. Summary of Tl-1223 samples: synthesis and annealing conditions, impurities, refined structural lattice parameters from XRD, and $T_c$ values from bulk magnetic susceptibility results.

| Sample # | Synthesis and annealing conditions | Impurities | *a* (Å) | *c* (Å) | Cu-O$_{planar}$ (Å) | Cu-O$_{apical}$ (Å) | V (Å$^3$) | c/a | $T_c$ (K) (magnetic) |
|---|---|---|---|---|---|---|---|---|---|
| 1 | as prepared | BaCuO$_2$ (7.7%), CaO (1.6%) | 3.84425(8) | 15.8871(1) | 1.923 | 2.765 | 234.78 | 4.1326 | 106 |
| 1-1 | 1+ annealing at 720°C in O$_2$ for 48 hours. | BaCuO$_2$ (4.9%), CaO (0.6%) | 3.84554(7) | 15.9066(3) | 1.923 | 2.768 | 235.22 | 4.1363 | 112 |
| 1-2 | 1-1 + annealing in flowing Ar+4%H$_2$ for 1 hour. | BaCuO$_2$ (3.6%), CaO (0.5%) | 3.84679(6) | 15.9199(3) | 1.924 | 2.770 | 235.57 | 4.1384 | 119 |
| 1-3 | 1-2 + annealing in flowing Ar+4%H$_2$ for 1 hour. | BaCuO$_2$ (4.5%), CaO (1.3%) | 3.84782(6) | 15.9246(3) | 1.925 | 2.771 | 235.77 | 4.1386 | 123 |
| 1-4 | 1-3 + annealing in flowing Ar+4%H$_2$ for 1 hour | BaCuO$_2$ (4.1%), CaO (3.0%) | 3.8480(8) | 15.9261(4) | 1.925 | 2.771 | 235.81 | 4.1387 | 125 |
| 2 | as prepared | Tl-1212 (7.6%), CaO (1.7%) | 3.8435(1) | 15.8801(5) | 1.923 | 2.763 | 234.58 | 4.1316 | 106 |
| 2-1 | 2 + annealing at 720°C in O$_2$ for 48 hours. | Tl-1212 (4.8%) CaO (1.8%) | 3.8443(1) | 15.8857(5) | 1.923 | 2.764 | 234.76 | 4.1322 | 110 |
| 2-2 | 2-1 + grinding + annealing at 675°C in N$_2$ for 1 hour. | BaCuO$_2$(10.6%), CaO (2.4%) | 3.84891(7) | 15.8755(3) | 1.925 | 2.762 | 235.18 | 4.1246 | 125 |
| 2-3 | 2-3 + pelletization + annealing in flowing N$_2$ for 1 hour. | BaCuO$_2$(10.6%), CaO (2.4%) | 3.84885(9) | 15.8755(4) | 1.925 | 2.762 | 235.17 | 4.1247 | 125 |

Figure 2 shows the (atomic number) Z-contrast STEM image of sample 1 oriented along the [100] axis. No intergrowths and supercell modulation are visible in this selected volume of the sample. The bright spots on the image are associated with heavier elements, such as Ba and Tl, while comparatively darker spots belong to lighter Ca and Cu; oxygen is not directly resolved in the images. At higher resolutions (Fig. 2b), the lattice can be observed to have a perfect elemental arrangement without any imperfection. STEM images for the final annealed samples 1-4 and 2-3 in Figure 3 show intensity variation in the Ca layer. Bright spots that are statistically distributed along with the Ca atomic layer can be designated to Tl$^{3+}$ which had been suggested in various structure refinements models.[4,15,26,27] In those reports it was concluded that Ca substitution by Tl at low concentration is always possible, but such substitutions need to be suppressed in order to reach a reasonably high $T_c$.[4] We however find that a higher $T_c$ is observed in samples with Tl substitutions at Ca site, even though the $T_c$ is not as high as HP synthesized sample.[28] No other atomic defects are visible, and at this point it is difficult to interpret reasons for the difference between the '*c*' lattice parameters of samples 1-4 and 2-3. The difference in the '*c*' lattice parameter may depend on the amounts of Tl$^{3+}$ atoms substituting Ca$^{2+}$, which are driven by the initial preparative and annealing conditions as we have discussed earlier.

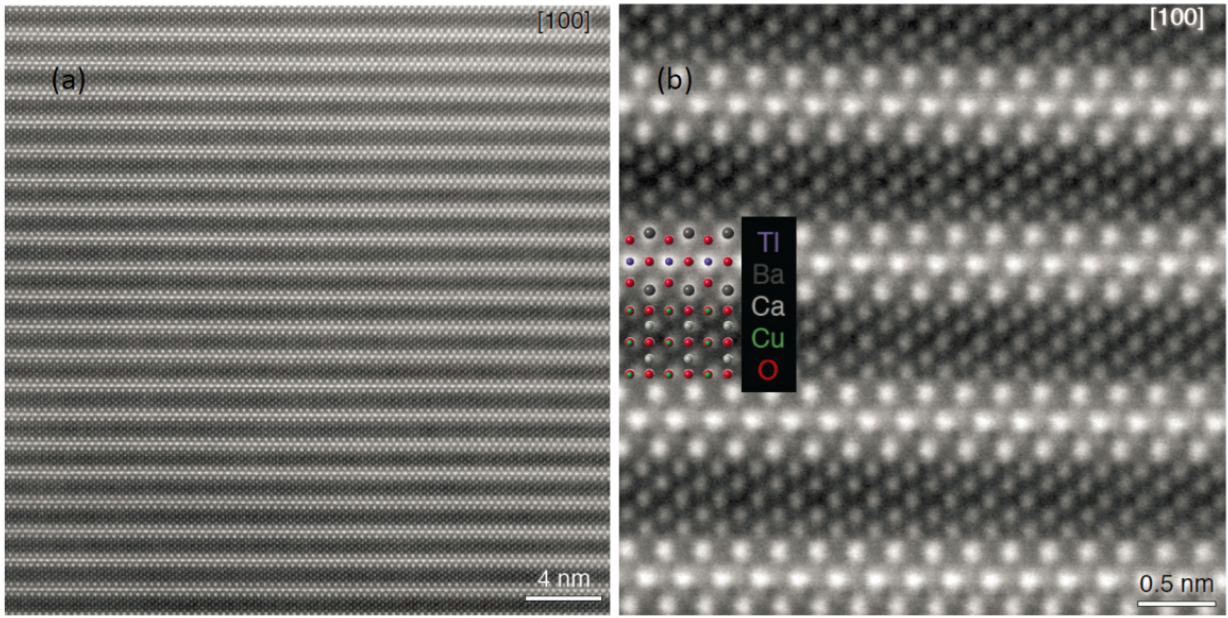

**Figure 2.** (a) Atomic resolution Z-contrast STEM image of a Tl-1223 grain projected along [100]. The lattice fringes show a periodicity for Tl-1223 structure, and no deviation from this periodicity is visible. (b) A magnified image of a small area shown in (a); a schematic model of atomic arrangement is shown in the left side of (b), where the Tl, Ba, Ca, Cu and O are represented by purple, Grey, white, green, and red circles, respectively (color online). No atomic defects are visible here.

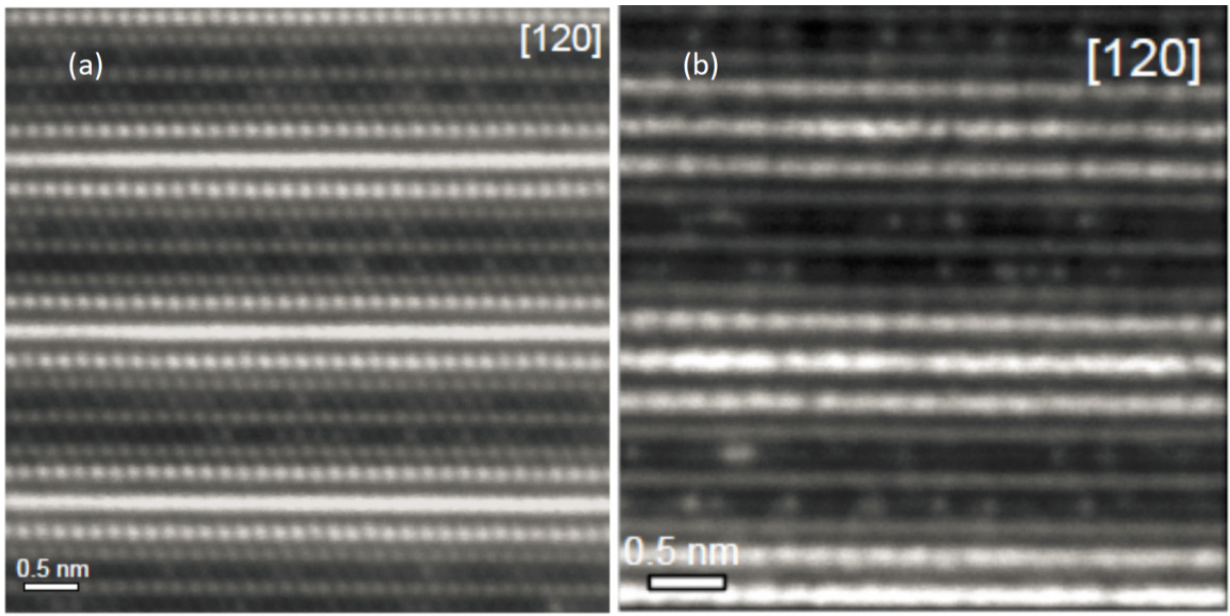

**Figure 3.** Atomic resolution Z-contrast STEM images of a Tl-1223 grain projected along [120] of the Ar+4%$H_2$ annealed sample 1-4 (a), and $N_2$ annealed sample 2-3 (b). Variations in the brightness of spots along the Ca layer indicate the presence atomic doping in this layer. The relative high intensity of some of the atomic columns in the Ca layer suggests that Tl is segregating into the Ca layer.

Figure 4 shows the ZFC and FC temperature-dependent magnetization of samples 1 and 2. There is a gradual increase in the onset of diamagnetic transition temperature from 106 K for sample 1, to 125 K for sample 1-4 (Fig. 4a). Similar change in $T_c$ is observed for sample 2 and $N_2$-annealed 2-3 (Fig. 4b). The ZFC curves of samples 1, 1-4 and 2 also show a second transition at around 100 K. This step is not present in sample 2-2, which was obtained after grinding, cold pressing, then annealing under flowing $N_2$ at 675 °C (2 hours). To clarify whether the low-temperature feature originates from the grain boundaries, grain alignment, or any other superconducting secondary phase, magnetic measurements were performed on powdered sample 1-4. A comparison between the ZFC-FC curves of sample 1-4 in the form of pellet and powder is shown (inset of Fig. 4a). The ZFC curve of powdered sample 1-4 shows single step with lowered screening and height almost equal to that of the first step of sample 1-4 (Fig. 4a). From the absence of second step in powdered samples, it can be speculated that other Tl-based phases are absent sample 1 and the second transition is mostly due to inter-granular current. Minute variation in the stoichiometry of grains may be possible as evitable from broad transition.

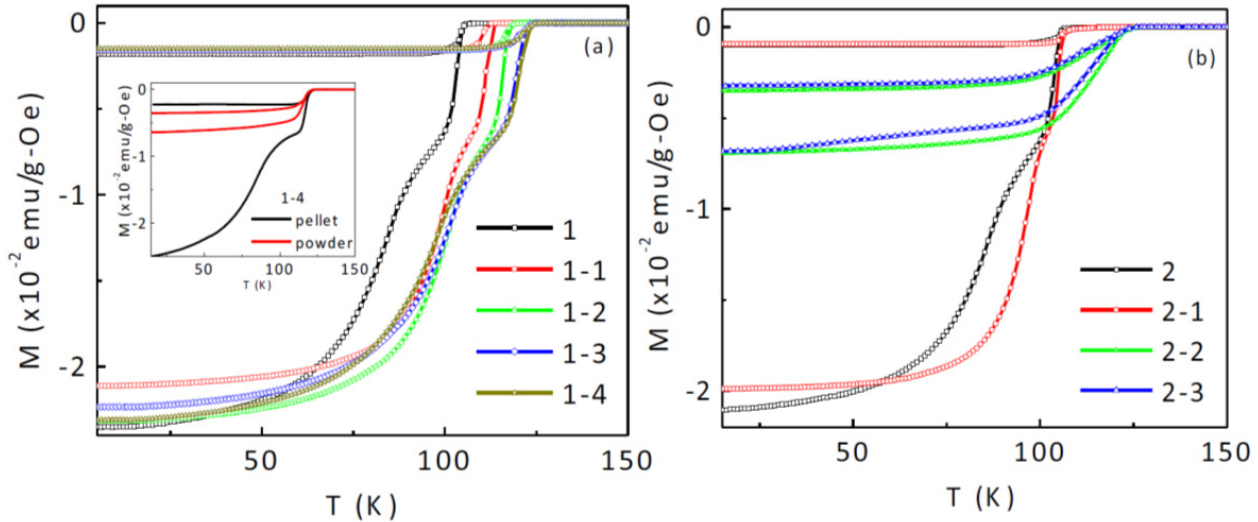

**Figure 4.** (a) Zero-field-cooled (ZFC) and field- cooled (FC) magnetic curves of the as-prepared, and annealed variations of Tl-1223 sample 1, measured under an applied field of 20 Oe. (b) The ZFC-FC curves of as-prepared sample 2, and the $O_2$ and $N_2$ annealed versions of it. Inset compares the ZFC-FC curve of Tl-1223 sample 1-4 in the form of pellet as well as powder. For listing of sample notations, see Table 1.

Figure 5 shows the temperature-dependent resistivity ($\rho$) and field-dependence of critical-current density ($J_c$) of a few as-prepared and annealed Tl-samples. Onset temperature for superconductivity increases with annealing (Fig. 5a). Normal state resistivity can be high; an order decrease in resistivity may be due to decrease in the amounts of impurities in sample 1-2. The normal state resistivity for sample 1-4 is high compared to sample 1-2 and can be related to the increase in the amount to impurities after the 3$^{rd}$ annealing. The superconducting resistive transition widths for all samples are comparable (~ 7 K). It is therefore clear that the superconducting properties starts to deteriorate after 3 hours annealing in Ar+4%$H_2$. For example, we also annealed another sample (not discussed here) for 12 hours and obtained a $T_c$=100 K, with decreased superconducting volume fractions, and more impurities. In order to estimate the critical current density ($J_c$), isothermal magnetic hysteresis curve were obtained at 5 K for sample 1-4 in both bulk and powder form. The $J_c$ of bulk sample in the form of rectangular bar shaped pellet was estimated using formula of $J_c = 20\Delta M/[a(1-a/3b)]$, where '$a$' and '$b$' are the lengths of the cross-section perpendicular to the applied field with $a < b$. For the powdered sample, the relation of $J_c = 30\Delta M/d$ was used, where $d$ (in $cm$) is average grain size estimated from scanning electron micrographs (example shown in inset of Fig. 5b). Here, $\Delta M$ (in emu/cm$^3$) is the width of hysteresis curve, which is almost same in both cases. $J_c$ values of ~ 8.65 x $10^4$ A/cm$^2$ and ~ 1 x $10^7$ A/cm$^2$ (Fig. 5b) were obtained at self-fields for the bulk pellet and powders, respectively. High $J_c$ for powder sample is similar to that reported in literature for polycrystalline samples.[29] This may be an over-estimated value because only the intra-granular component of $J_c$ is considered, not a good model for superconducting grains that are not well separated.

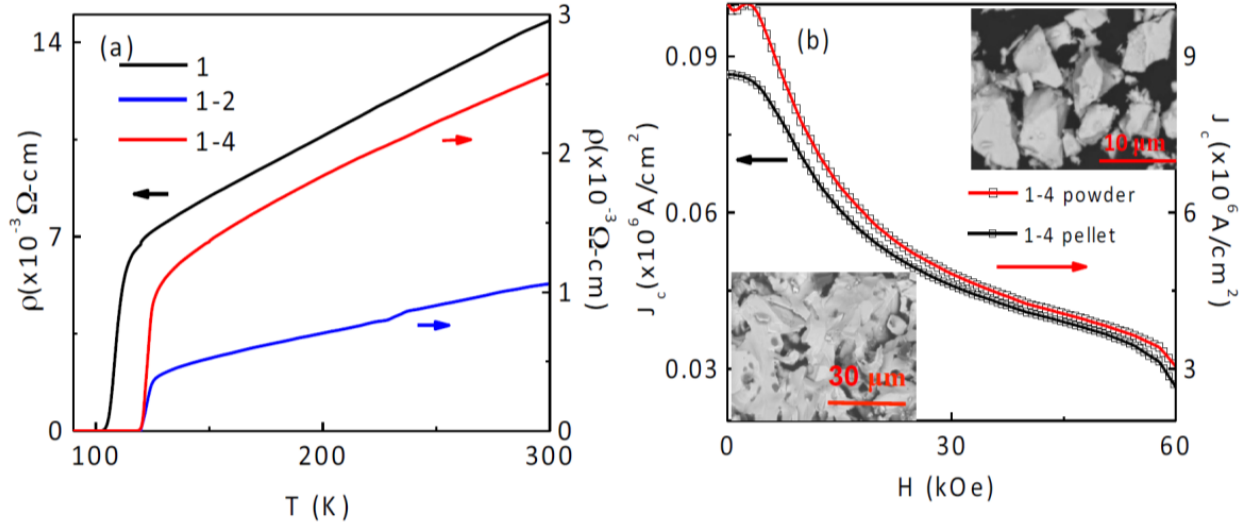

**Figure 5**. Resistivity ($\rho$) versus temperature ($T$) plots of as-prepared and annealed sample 1 is shown in figure (a). Comparison between the critical current densities of pellet and powdered Tl-1223 sample 1-4, measured at 5 K. Upper and lower insets shows the scanning electron micrographs of sample 1-4 in the form of powder and pellet.

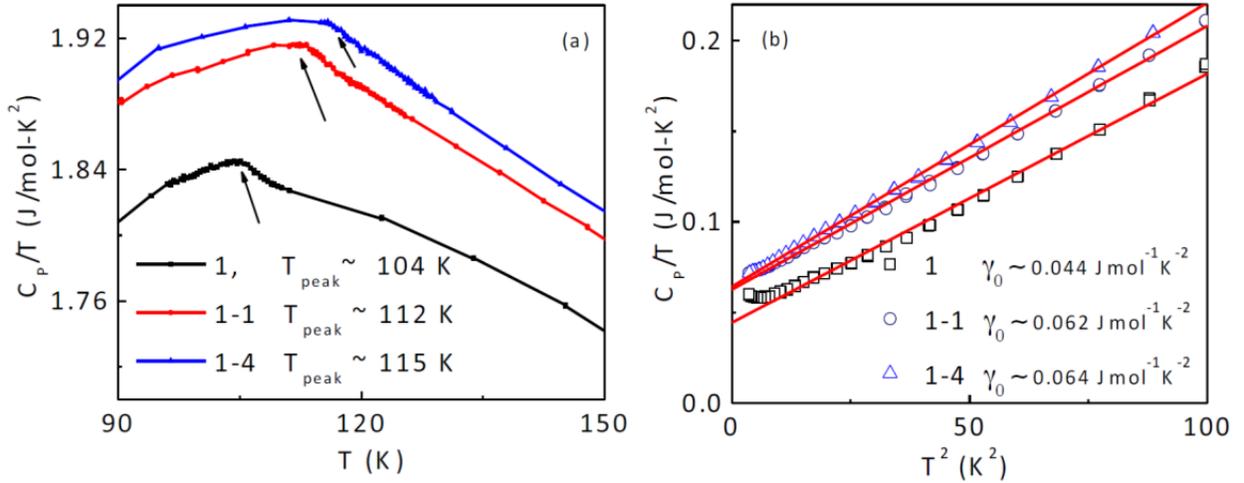

**Figure 6**. Specific heat capacity ($C_P$) of the as-prepared and annealed Tl-1223, plotted as $C_P/T$ vs, $T$ near $T_c$ (a), and as $C_P/T$ vs, $T^2$ below 10 K with linear fits (b).

Figure 6 shows the temperature-dependent specific heat capacity ($C_P$) of the as-prepared and annealed sample near $T_c$ (Fig. 6a) and below 10 K (Fig. 6b). Arrows indicate a small protuberance near the superconducting $T_c$ of the samples. The peak for sample 1 is well resolved at $T$ ~ 106 K, same as the diamagnetic onset temperature. With annealing the peak shifts to a higher temperature but also gets broaden. Low temperature data was fitted below 10 K according to the expression $C_P/T = \gamma_0 + \beta T^2$. The residual linear Sommerfeld electronic term, $\gamma_0 = 0.044$ J.mol$^{-1}$K$^{-2}$ for the as-prepared sample 1, which actually increases to 0.062 J.mol$^{-1}$K$^{-2}$ and 0.064 J.mol$^{-1}$K$^{-2}$ in annealed samples 1-1, and 1-4, respectively. High values for $\gamma_0$ may appear because of increased disorder, phase segregation, impurities, etc. Below 3 K we observe a small upturn in the as-prepared sample, which gets suppressed after annealing. The upturn is reported to be associated with free Cu$^{+2}$ moments that order at low temperature and either become a Schottky anomaly at high fields or may give rise to spin-glassy behavior.[30] We must note that for sample 2-3, no peak in heat capacity is visible near the $T_c$.

**Conclusions**

An efficient way of preparing bulk and polycrystalline Tl-1223 superconductor is reported here at ambient pressures. Use of excess Tl$_2$O$_3$ is important to get almost a single phase of the compound. Short time repeated grinding and heating ensured better homogeneity and helps in decreasing secondary phases. As-prepared samples had a lower $T_c$, which increased by 18% after short time thermal annealing in a flowing gas mixture of Ar+4%H$_2$ or N$_2$ gases. We found substitution of Ca by Tl in the final annealed samples, while no such imperfection is observed in the as-prepared sample.

**Acknowledgments**

This research was supported by the National Science Foundation (NSF) through grant number of DMR-0938330 (RS), and also the Center for Nanophase Materials Sciences (CNMS), which is sponsored at ORNL by the Scientific User Facilities Division, Office of Basic Energy Sciences, U.S. DOE (JI). In addition, the partial work was carried out as part of funding through U.S. Department of Energy, Office of Science, Basic Energy Sciences, Materials Science and Engineering Division (AS and RS).